\documentclass[aps,prd,10pt,superscriptaddress,twocolumn,nofootinbib]{revtex4-1}
\usepackage{graphicx, epsfig} 
\usepackage{amsmath,amssymb,amsfonts,dsfont,mathrsfs,amsthm,mathtools}
\usepackage{bm} 
\usepackage{color}
\usepackage[usenames]{xcolor}
\usepackage{hyperref}
\usepackage{siunitx}
\hypersetup{colorlinks=true,urlcolor=blue,linkcolor=magenta,citecolor=blue,filecolor=blue}
\usepackage[normalem]{ulem}
\usepackage{array}
\usepackage{hyperref}
\usepackage{booktabs}
\usepackage{blindtext}
\usepackage{slashed}

\newcommand{\diff}[1]{\text{d}#1}

\newcommand{\Lag}{\mathscr{L}}

\begin{document}

\title{Axial anomaly in nonlinear conformal electrodynamics}

\author{Francisco Colip\'i-Marchant}
\email{francisco.colipi@ug.uchile.cl}
\affiliation{Departamento de Física, Facultad de Ciencias Físicas y Matemáticas, Universidad de Chile, Blanco Encalada 2008, Santiago, Chile}

\author{Crist\'obal Corral}
\email{crcorral@unap.cl}
\affiliation{Instituto de Ciencias Exactas y Naturales, Universidad Arturo Prat, Avenida Playa Brava 3256, 1111346, Iquique, Chile}
\affiliation{Facultad de Ciencias, Universidad Arturo Prat, Avenida Arturo Prat Chac\'on 2120, 1110939, Iquique, Chile}

\author{Daniel Flores-Alfonso}
\email{danflores@unap.cl}
\affiliation{Instituto de Ciencias Exactas y Naturales, Universidad Arturo Prat, Avenida Playa Brava 3256, 1111346, Iquique, Chile}
\affiliation{Facultad de Ciencias, Universidad Arturo Prat, Avenida Arturo Prat Chac\'on 2120, 1110939, Iquique, Chile}

\author{Leonardo Sanhueza}
\email{lsanhueza@udec.cl}
\affiliation{Departamento de F\'isica, Universidad de Concepci\'on, Casilla, 160-C, Concepci\'on, Chile}

\begin{abstract}
We study the axial anomaly of Dirac spinors on gravitational instanton backgrounds in the context of nonlinear electrodynamics. In order to do so, we consider Einstein gravity minimally coupled to a recently proposed conformal electrodynamics that enjoys duality transformation invariance. These symmetries allow us to generalize the Eguchi-Hanson configuration while preserving its geometry. We then compute the Dirac index of the nonlinearly charged Eguchi-Hanson and Taub-NUT configurations. We find that there is an excess of positive chiral Dirac fermions over the negative ones which triggers the anomaly. 
\end{abstract}

\maketitle

\section{Introduction}

The axial anomaly is a remarkable phenomenon that occurs when the axial symmetry of a classical system is broken due to quantum effects, providing one of the most interesting examples where the nontrivial topology of fiber bundles has direct and measurable consequences in physics. It has been observed through the anomalous decay $\pi^0\to\gamma\gamma$ in the standard model, which would have been otherwise forbidden at the classical level since a two-photon state does not couple to an axial current~\cite{Steinberger:1949wx,10.1143/ptp/4.3.347,Schwinger:1951nm,Adler:1969gk,Bell:1969ts,Bardeen:1969md}. On the other hand, in condensed matter systems, the axial anomaly is a universal phenomenon that appears in three-dimensional metals in presence of electric and magnetic fields, producing a longitudinal magnetoresistance which is absent in the classical theory of magnetotransport~\cite{Goswami:2015uxa}. Additionally, it has been shown that the topological transport and the chiral magnetic effect can be understood in terms of the axial anomaly in Weyl semimetals~\cite{Zyuzin:2012tv}. Indeed, the latter has been detected in zirconium pentatelluride~\cite{Li:2014bha} and particle colliders have presented experimental evidence in favor of their existence by analyzing the quark-gluon plasma~\cite{Belmont:2014lta,STAR:2015wza}. In topological insulators~\cite{1985ZhPmR..42..145V,PANKRATOV198793,doi:10.1126/science.1148047,PhysRevLett.95.226801,PhysRevLett.96.106802}, the effective field theory describing the electromagnetic responses are governed by the axial anomaly~\cite{2012PhRvB..85d5104R}. These materials have exotic properties such as spin-momentum locking~\cite{2009Natur.460.1101H}, the presence of Majorana states in the superconducting phase~\cite{PhysRevLett.100.096407}, and the quantum spin hall effect~\cite{PhysRevLett.95.226801}; crucial features for spintronic devices and dissipationless transistors in quantum computing.

The origin of the axial anomaly can be traced back to the imbalance of the number of spinors with different chirality, produced by their interaction with gauge fields possessing nontrivial topology; this usually happens in presence of instantons~\cite{Jackiw:1976dw,Jackiw:1977pu}. For instance, the Atiyah-Patodi-Singer theorem~\cite{Atiyah:1968mp,APS-eta} provides a deep connection between two topological indices of differential operators acting on spinors, giving a concrete prescription to know if the background fields would break the axial symmetry. This fact led Fujikawa to conclude that the axial anomaly is a nonperturbative effect that arises from the nontrivial transformation of the fermionic measure in the path-integral formulation of gauge theories~\cite{Fujikawa:1979ay,Fujikawa:1980eg}. 

Charap and Duff noticed that when gauge theories are coupled to gravity, the topology of the matter content may be dictated by the background~\cite{Charap:1977ww}. Moreover, different instantons exist on spacetimes with inequivalent topologies. In particular, there are solutions of the Yang-Mills equations which cannot exist in flat space or would otherwise be singular~\cite{Charap:1977re}. These conclusions were reached by analyzing (anti-)self-dual fields on Euclidean space. In such a scenario, the energy-momentum tensor vanishes by virtue of the matter's conformal symmetry in addition to (anti-)self duality. Their lack of backreaction on these backgrounds renders their gravitational effect quite subtle. Some physical aspects that involve nontrivial gravitational and matter fields are even less obvious, such is the case of the axial anomaly as showcased by Pope in Refs.~\cite{Pope:1978zx,Pope:1981jx}.

In this paper, we study the axial anomaly in electromagnetism beyond Maxwell's theory. On the one hand, it is well-known that the effective theory of quantum electrodynamics induces nonlinear effects when the fermionic degrees of freedom are integrated out, capturing quantum effects at one-loop level produced by virtual fermions~\cite{Heisenberg:1936nmg}. On the other hand, nonlinear constitutive relations have also been constructed to render the self-energy of charged point particles finite~\cite{Born:1933qff,Born:1934ji,Born:1934gh}. Moreover, when coupled to gravity, nonlinear electrodynamics has been considered an interesting framework to study black holes~\cite{Ayon-Beato:1998hmi,Bronnikov:2000vy,Ayon-Beato:2000mjt,Cataldo:2000we,Burinskii:2002pz,Hassaine:2007py,Miskovic:2010ey,Cisterna:2020rkc,Bravo-Gaete:2022mnr,DiazGarcia:2022jpc,Ayon-Beato:2022dwg,Bravo-Gaete:2022mnr,Alvarez:2022upr}, conserved charges~\cite{Miskovic:2010ui}, holographic superconductors~\cite{Jing:2011vz,Aranguiz:2013cwa} and black strings~\cite{Cisterna:2021ckn}, among others~\cite{Babaei-Aghbolagh:2022uij,Babaei-Aghbolagh:2022itg,Ferko:2022iru,Arenas-Henriquez:2022ntz,Ferko:2023ruw}. A standard reference is Pleba\'nski's work~\cite{Plebanski:1970}, while a recent review is given in Ref.~\cite{Sorokin:2021tge}.

Given that the nature of the axial anomaly has proved to be fine-drawn, in this work, we opt to carry out our exploration on (anti-)self-dual fields for a conformal nonlinear electrodynamics which, by definition, satisfy the electrodynamic equations of motion automatically. However, a duality transformation does not necessarily leave the constitutive relation of the theory invariant. There are, however, ways to test if a given theory is invariant under duality rotations, see for example Refs.~\cite{Gaillard:1981rj,Bialynicki-Birula:1983}. As it turns out, if one imposes duality rotation invariance on arbitrary nonlinear electrodynamics it can be concluded that there are infinitely many of them, as many as real functions of a single variable~\cite{Salazar:1987ap,Gibbons:1995cv}. 
Among this class of theories is the recently proposed one-parameter family of Lagrangians dubbed \emph{ModMax}~\cite{Bandos:2020jsw}. This constitutes the only nonlinear extension of Maxwell's equations that is invariant under both duality and conformal transformations. The only other theory known to possess such symmetries is the strictly non-linear Bia{\l}ynicki-Birula electrodynamics whose Lagrangian vanishes as a consequence of its constraints; for further details see Ref.~\cite{Bialynicki-Birula:1992rcm}. Not only is that theory conformal and invariant under duality rotations but also under a larger SL(2,$\mathbb{R}$) duality transformation group.

Very recently, explicit Lagrangians for all duality-symmetric nonlinear electrodynamics were derived in Ref.~\cite{Avetisyan:2021heg} using a democratic formulation where duality invariance is manifest. This approach has been further discussed in Ref.~\cite{Avetisyan:2022zza}. We mention that these works were built on the democratic
formulation of the free Maxwell theory established in Ref.~\cite{Mkrtchyan:2019opf}; for earlier work on the subject see Ref.\cite{Pasti:1995ii}. Outside of this setting, ModMax
has been studied in the context of black holes, birefringence, supersymmetry, gravitational instantons, wormholes and the 't Hooft anomaly, among other applications~\cite{Flores-Alfonso:2020euz,Amirabi:2020mzv,Bandos:2021rqy,Chatzistavrakidis:2021dqg,Barrientos:2022bzm,BallonBordo:2020jtw,Flores-Alfonso:2020nnd}.

In this work, we show that gravitational instantons coupled to (anti-)self-dual fields in nonlinear electrodynamics trigger the axial anomaly of charged Dirac spinors. This represents the first concrete example of the axial anomaly in ModMax theory. To this end, we study the field equations of the Einstein-ModMax system when the background has an underlying nontrivial circle fibration over a K\"ahler manifold. Such is the case of the Taub-NUT and Eguchi-Hanson spaces. We begin reviewing the former case as it has already been discussed in the literature~\cite{BallonBordo:2020jtw,Flores-Alfonso:2020nnd}. We then formulate an ansatz of Eguchi-Hanson type and find a new self-gravitating ModMax configuration. Both the electromagnetic and gravitational curvatures satisfy (non-linear) self-duality relations. The difference between positive and negative chiral charged fermions is computed through the Atiyah-Patodi-Singer theorem taking into account the boundary contribution. Particular attention is paid to the computation of the $\eta_D$-invariant that measures the difference between positive and negative eigenvalues of the Dirac operator evaluated on the tangential components of the boundary. We find that the nonlinearly charged Eguchi-Hanson and Taub-NUT instantons induce an imbalance between Dirac spinors of different chirality, triggering the anomaly. We comment on possible applications regarding other types of anomalies. 

The article is organized as follows: in Sec.~\ref{sec:modmax}, we briefly review the nonlinear electrodynamics that we shall couple to Einstein gravity. The Einstein-ModMax system constitute our main theoretical setup. Then, in Sec.~\ref{sec:instantons}, we study gravitational instantons as configurations that solve these equations. We analyze their properties paying close attention to (anti-)self-duality. Section~\ref{sec:index} is devoted to computing the index of the Dirac operator in presence of nonlinear electromagnetic fields. Final comments and discussions are given in Sec.~\ref{sec:discussion}.

\section{Nonlinear electrodynamics\label{sec:modmax}}

Departing from the superposition principle of electromagnetic fields leads to a generalization of the source-free Maxwell's equations given by
\begin{align}
    \nabla_\mu P^{\mu\nu} = 0\;\;\;\;\; \mbox{and} \;\;\;\;\; \nabla_\mu \tilde{F}^{\mu\nu}=0 \,, \label{eoma}
\end{align}
where $P^{\mu\nu}=P^{\mu\nu}(F^{\alpha\beta})$ has a nonlinear dependence on the Maxwell's field strength and $\tilde{F}_{\mu\nu}=\tfrac{1}{2}\varepsilon_{\mu\nu\lambda\rho}F^{\lambda\rho}$. The nonlinear electrodynamics dubbed ModMax is the unique nonlinear extension of Maxwell theory that preserves both conformal symmetry and $SO(2)$ duality invariance~\cite{Bandos:2020jsw}. Originally the theory was formulated in a Legendre dual formulation and its Lagrangian was derived afterwards, However, a straightforward derivation has been given in Ref.~\cite{Kosyakov:2020wxv}. An additional symmetry of ModMax theory is that it is Legendre self-dual~\cite{Flores-Alfonso:2020euz}, similar to Born-Infeld theory~\cite{Gibbons:2000xe}. In this sense, both electrodynamics are somewhat similar, however, the Born-Infeld case is among the few theories where birefringence is absent~\cite{Russo:2022qvz}. Indeed, the ModMax electrodynamics exhibits birefringence of electromagnetic waves when the field equations are linearized around non-vacuum configurations of constant electromagnetic fields~\cite{Bandos:2020jsw}. Nevertheless, the theory is continuously connected to Maxwell electrodynamics in the linear limit.

It should be kept in mind that further below we use ModMax theory as the matter source of the Einstein equations. Thus, we find it convenient to formulate it using the following action principle
\begin{align}\label{IModMax}
    I_{\rm M} = -\int_{\mathcal{M}}\diff{^4x}\sqrt{|g|}\,\Lag_{\rm M}\,,
\end{align}
where $g=\det g_{\mu\nu}$ is the determinant of the metric and the Lagrangian is given by 
\begin{align}\label{LagModMax}
    \Lag_{\rm M} = X\cosh\gamma - \sqrt{X^2-Y^2}\,\sinh\gamma\,.
\end{align}
Here, $X=\tfrac{1}{4}F_{\mu\nu}F^{\mu\nu}$ and $Y=\tfrac{1}{4}\tilde{F}_{\mu\nu}F^{\mu\nu}$ are the two standard independent Euclidean invariants that can be constructed with the field strength and its dual.\footnote{In the Lorentzian version of the theory one should change $Y^2\to-Y^2$ in Eq.~\eqref{IModMax} and variations thereof.} Notice that $X$ has even parity, while, $Y$ is odd. Thus, ModMax theory is a parity preserving nonlinear extension of Maxwell's; yet another similarity with the Born-Infeld case. Additionally, the parameter $\gamma$ measures the degree of nonlinearity of the electromagnetic fields. When $\gamma=0$ the theory becomes the standard linear Maxwell electrodynamics. Notice that the parameter is dimensionless, as required by conformal symmetry. In the Lorentzian formulation, causality constraints impose that $\gamma\geq0$~\cite{Bandos:2020jsw}.

As usual, the Bianchi identity allows for the gauge potential $A=A_\mu\diff{x^\mu}$ to be defined by $F_{\mu\nu}=\partial_\mu A_\nu-\partial_\nu A_\mu$. Moreover, by performing arbitrary variations of the action~\eqref{IModMax} with respect to $A_\mu$, yields Eq. \eqref{eoma} with the constitutive relation
\begin{align}\label{PModMax}
    P_{\mu\nu} &= \left(\cosh\gamma - \frac{X\,\sinh\gamma}{\sqrt{X^2-Y^2}} \right)F_{\mu\nu} + \frac{Y\sinh\gamma}{\sqrt{X^2-Y^2}}\,\tilde{F}_{\mu\nu}\,.
\end{align}
A straightforward calculation shows that the latter satisfies $\tilde{F}_{\mu\nu}F^{\mu\nu}=\tilde{P}_{\mu\nu}P^{\mu\nu}$, a sufficient condition for it to be invariance under duality rotations as shown in Ref.~\cite{Gibbons:1995cv}. The stress-energy tensor, on the other hand, is defined as the variation of the ModMax action with respect to the metric; in particular, using the definitions in Eqs.~\eqref{LagModMax} and \eqref{PModMax}, it gives
\begin{align}\label{TmunuModMax}
     T_{\mu\nu} = P_{(\mu }^{\ \ \lambda}F_{\nu) \, \lambda} - g_{\mu\nu}\,\Lag_{\rm M}\,.
\end{align}
Notice that the latter reduces to the standard Maxwell stress-energy tensor in the limit $\gamma\to0$. Since the theory is conformal, the trace of Eq.~\eqref{TmunuModMax} vanishes. Whenever this condition is broken at the quantum level it is known as the trace anomaly~\cite{Capper:1974ic,Deser:1976yx,Duff:1977ay}.

To couple the theory with the Einstein-Hilbert action, we consider
\begin{align}
    I[g_{\mu\nu},A_\mu] = \kappa\int_{\mathcal{M}}\diff{^4x}\sqrt{|g|}\left(R-2\Lambda \right) + I_{\rm M}\,,
\end{align}
where $\kappa=(16\pi G)^{-1}$ with $G$ being Newton's constant, $\Lambda$ is the cosmological constant and $R=g^{\mu\nu}R^{\lambda}{}_{\mu\lambda\nu}$ is the Ricci scalar. In this case, the field equations for the metric are
\begin{align}\label{eomg}
    R_{\mu\nu} - \frac{1}{2}g_{\mu\nu}R + \Lambda g_{\mu\nu} = \frac{1}{2\kappa}T_{\mu\nu}\,.
\end{align}
Taking the trace on the latter and using the fact that $T^\mu_\mu=0$ one obtains $R=4\Lambda$. Therefore, all solutions to the field equations of Einstein-ModMax theory have constant Ricci scalar and Eq.~\eqref{eomg} reduces to $R_{\mu\nu}=\Lambda g_{\mu\nu}+8\pi G T_{\mu\nu}$. Thus, the system of second-order non-linear partial differential equations given by~\eqref{eoma} and~\eqref{eomg} determines the dynamics of the theory.

\section{(Anti-)self-dual instantons\label{sec:instantons}}

Central to our calculation of the Dirac index are systems with non-trivial Pontryagin invariants. Indeed, whenever the electric and magnetic components of the Weyl tensor are orthogonal, the Pontryagin density vanishes. Thus backgrounds with (anti-)self-dual curvature are of particular interest to us. In this section, we study nonlinearly charged self-gravitating systems with a nonvanishing Pontryagin term. These configurations solve Einstein's field equations~\eqref{eomg} and are described geometrically as inhomogeneous metrics on complex line bundles over K\"ahler manifolds~\cite{Page:1985bq}. The gauge potential is aligned along the fiber which allows the eletromagnetic fields to simultaneously satisfy the ModMax constitutive relation~\eqref{PModMax} and field equations~\eqref{eoma}.

In particular, we focus on metrics of the Taub-NUT~\cite{Taub:1950ez,Newman:1963yy} and Eguchi-Hanson~\cite{Eguchi:1978xp,Eguchi:1978gw} class. These Euclidean spaces are known as gravitational instantons~\cite{Hawking:1976jb,Eguchi:1976db,Gibbons:1978tef,Hawking:1978ghb,Gibbons:1979xm,Eguchi:1978xp,Eguchi:1978gw,Eguchi:1979yx,Eguchi:1980jx} due to their close resemblance with Yang-Mills pseudo-particle configurations.

\subsection{Taub-NUT}

The Lorentzian Taub-NUT spacetime is a one-parameter generalization of the Schwarzschild black hole that is usually interpreted as a gravitational dyon, in the sense that it is endowed with an electric and magnetic mass; see for instance Refs.~\cite{Lynden-Bell:1996dpw,Araneda:2016iiy}. They source the electric and magnetic components of the Weyl tensor, respectively. In Euclidean signature, metrics of this type are described by
\begin{align}\label{taubnutansatz}
    \diff{s^2} = f(r)\left(\diff{\psi} + 2n\mathcal{B}_{(k)} \right)^2 + \frac{\diff{r^2}}{f(r)} + \left(r^2-n^2\right)\diff{\Sigma^2_{(k)}}\,,
\end{align}
where $f(r)$ is determined from the field equations. Here, $\mathcal{B}_{(k)}$ is the K\"ahler potential $1$-form that defines the Hermitian symplectic exact form $\Omega_{(k)}=\diff{\mathcal{B}_{(k)}}$ associated to the Einstein-K\"ahler base manifold of constant curvature $\diff{\Sigma_{(k)}^2}$. For $k=1,0,-1$, the base manifold is topologically $\mathbb{S}^2$, $\mathbb{T}^2$, and $\mathbb{H}^2$, respectively, and their K\"ahler potential and line element can be parametrized as shown in Table~\ref{Table}.
\begin{center}
\begin{table}[h!]
\begin{tabular}{cccc}
\cline{1-4}
 & $k=1$ \,  &  $k=0$ \, & $k=-1$ \, \\ 
 \hline
 $\mathcal{B}_{(k)}$ \,  & $\cos\vartheta\diff{\varphi}$ \, & $\vartheta\diff{\varphi}$ \, & $\cosh\vartheta\diff{\varphi}$  \\ 
 \hline
  $\diff{\Sigma_{(k)}^2}$ \, & $\diff{\vartheta^2} + \sin^2\vartheta\diff{\varphi^2}$ \, & $\diff{\vartheta^2}+\diff{\varphi^2}$ \,
  & $\diff{\vartheta^2} + \sinh^2\vartheta\diff{\varphi^2}$ \\
 \hline
\end{tabular}
\caption{\label{Table}Line element of the transverse sections and K\"ahler potential $1$-form for $\mathbb{S}^2$, $\mathbb{T}^2$, and $\mathbb{H}^2$ topologies, respectively.}
\end{table}
\end{center}
The first charged metric in the Taub-NUT class was found by Brill in Ref.~\cite{Brill:1964tn}. More recently, higher-dimensional versions have been constructed in Ref.~\cite{Awad:2005ff}. In these configurations, the Maxwell field is assumed to be proportional to their associated K\"ahler potential $1$-form, namely,
\begin{align}\label{ATNansatz}
    A=A_\mu\diff{x^\mu} = \alpha(r)\left(\diff{\psi} + 2n\mathcal{B}_{(k)} \right)\,.
\end{align}
From hereon, we shall assume the same structure for the metric and Maxwell fields as those given in Eqs.~\eqref{taubnutansatz} and~\eqref{ATNansatz}, respectively.

Inserting these ans\"atze into the field equations~\eqref{eoma} and~\eqref{eomg}, the linearly independent ordinary differential equations are  
\begin{align}
\label{eomatnut}
    0&= \alpha'' + \frac{2r\alpha'}{\left(r^2-n^2 \right)} - \frac{4n^2e^{-2\gamma}\alpha}{\left(r^2-n^2 \right)^2}\,,\\
    \notag
    0&=4\kappa\left[rf'+\left(\frac{r^2+n^2}{r^2-n^2}\right)f +\Lambda\left(r^2-n^2\right) - k\right] \\
    \label{eomgttnut}
    & - \left(r^2-n^2 \right)e^{\gamma}\alpha'^2+\frac{4n^2e^{-\gamma }\alpha^2}{\left(r^2-n^2 \right)} \,,
\end{align}
respectively, where $\alpha=\alpha(r)$ and $f=f(r)$, with prime denoting differentiation with respect to the radial coordinate. The system is solved by integrating first Eq.~\eqref{eomatnut} and then replacing its solution into Eq.~\eqref{eomgttnut} to solve for the metric function; this yields
\begin{subequations}\label{solnut}
\begin{align}\notag
    f(r) &= k\left(\frac{r^2+n^2}{r^2-n^2} \right) - \frac{2mGr}{r^2-n^2} - \frac{e^{-\gamma}}{4\kappa}\left(\frac{q^2-p^2}{r^2-n^2} \right) \\
    &- \frac{\Lambda}{3}\left(\frac{r^4-6n^2r^2-3n^4}{r^2-n^2}\right) \,,\\
    \alpha(r) &= \frac{1}{2n}\left[q\sinh\Phi(r) + p\cosh\Phi(r) \right]\,,
\end{align}
\end{subequations}
where $m$, $q$, and $p$ are integration constants and we have defined
\begin{align}
    \Phi(r) = e^{-\gamma}\ln\left(\frac{r-n}{r+n} \right)\,.
\end{align}
This solution was first found in Refs.~\cite{BallonBordo:2020jtw,Flores-Alfonso:2020nnd} and it is generically endowed with a conical defect wherever the metric function $f$ vanishes. Moreover, when $k=1$ an additional defect is present at $\vartheta=0$ and $\vartheta=\pi$ known as a Misner string~\cite{Misner:1963fr}. In Lorentzian signature, the latter may lead to closed time-like curves. Due to this, it has been extensively studied in Euclidean signature; where it bears a close resemblance with Yang-Mills instantons. Remarkably, it has been established that the presence of the Misner string contributes to the entropy in a nontrivial way~\cite{Mann:1999pc,Mann:1999bt,Hawking:1998ct,Mann:2004mi,Ciambelli:2020qny,Garfinkle:2000ms}. The reason for this is that it represents an obstruction to the foliation of the space with a scalar function that represents the Hamiltonian evolution of the system~\cite{Hawking:1998jf}. Nevertheless, recently the Lorentzian case has attracted a lot of interest since it has been shown that the Misner string is transparent for geodesic observers~\cite{Clement:2015cxa}. This, in turn, has lead to a new thermodynamic treatment of the spacetime, where the nut charge behaves as a genuine thermodynamic variable with its respective conjugated pair, enlarging the phase space of the system~\cite{Durka:2019ajz,Hennigar:2019ive,Bordo:2019slw,BallonBordo:2019vrn,Frodden:2021ces}.

Let us focus on the case when $k=1$. First, notice that by performing the coordinate transformation $\psi\to\tfrac{\psi}{2n}$, the line element~\eqref{taubnutansatz} can be written as
\begin{align}\label{taubnutselfdual}
    \diff{s^2} &= \frac{\diff{r^2}}{f(r)} + \left( r^2-n^2\right)\left[\sigma_1^2+\sigma_2^2 + \frac{4n^2f(r)\sigma_3^2}{\left(r^2-n^2 \right)} \right]\,,
\end{align}
where $\sigma_i$ are the left-invariant Maurer-Cartan one-forms of $SU(2)$ satisfying $\diff{\sigma_i}+\tfrac{1}{2}\epsilon_{ijk}\sigma_j\wedge\sigma_k=0$. Using the basis of Euler angles with $0\leq\vartheta\leq\pi$, $0\leq\varphi\leq 2\pi$ and $0\leq\psi\leq \beta_\psi$, they can be explicitly represented as
\begin{subequations}\label{MCformsu2}
\begin{align}
\sigma_1 &= \cos\psi \,\diff{\vartheta} + \sin\vartheta \, \sin\psi \, \diff{\varphi}\,, \\
\sigma_2 &= -\sin\psi\, \diff{\vartheta} + \sin\vartheta\,  \cos\psi\,  \diff{\varphi}\,, \\
\sigma_3 &= \diff{\psi} +\cos \vartheta\, \diff{\varphi}\, . 
\end{align}
\end{subequations}
In presence of a negative cosmological constant, say $\Lambda=-\tfrac{3}{\ell^2}$, we find that, if the conditions
\begin{align}
    p = \pm q \;\;\;\;\; \mbox{and} \;\;\;\;\;
    mG = 1-\frac{4 n^2}{\ell^2} \,,
\end{align}
are met, the metric~\eqref{taubnutselfdual} possesses an (anti-)self-dual Weyl tensor, i.e. $W_{\mu\nu\lambda\rho}=\pm\tilde{W}_{\mu\nu\lambda\rho}$, and the Killing vector field $\partial_\psi$ acquires a zero-dimensional set of fixed points at $r=n$. Then, the solution becomes
\begin{align}\label{fnut}
    f_{\rm nut}(r) &= \frac{r-n}{r+n} + \frac{\left(r-n \right)^2\left(r+3n \right)}{\ell^2(r+n)}\,,\\
    \alpha_{\rm nut}(r) &= \frac{q}{2n}\left(\frac{r-n}{r+n}\right)^{e^{-\gamma}}\,.
\end{align}
This is the (anti-)self-dual Taub-NUT-AdS space coupled to a ModMax field. Notice that the latter does not backreact on the metric since $T_{\mu\nu}$ vanishes identically, even though the field strength is nontrivial. This is a  unique feature of conformal theories whenever the fields are (anti-)self dual. In nonlinear electrodynamics, the latter condition is described by
\begin{align}\label{selfdualmodmax}
    F_{\mu\nu} = \pm \tilde{P}_{\mu\nu}\,.
\end{align}
In this case, the range of coordinates is $n\leq r<\infty$, $0\leq\vartheta\leq\pi$ and $0\leq\varphi\leq2\pi$. The absence of conical singularities, in turn, implies that $0\leq\psi\leq4\pi $. Then, this metric is topologically equivalent to $\mathbb{R}^4$ near the zero-dimensional set of fixed points. Additionally, in the limit $\gamma\to0$, one recovers the charged Euclidean (anti-)self-dual Taub-NUT solution of Ref.~\cite{Brill:1964tn}. Moreover, the ModMax field is asymptotically Maxwell, e.g., $\alpha(r)\to q/(2n)$ as $r\to\infty$. Thus, although the nonlinearly charged ModMax Taub-NUT solution is locally inequivalent to the Einstein-Maxwell one, they have the same asymptotic behavior.

Another instanton within the same class as discussed above is the one called Taub-Bolt. It is obtained when the Killing vector field $\partial_\psi$ has a two-dimensional set of fixed points at $r=r_b$, with $r_b>n$, defined as the largest root of the polynomial $f(r_b)=0$. Such set of fixed points are known as bolts, in contrast to the zero-dimensional variety which are called nuts. The ModMax case has been studied in detail before but not its contribution to the axial anomaly. Here, however, we shall focus on the globally (anti-)self-dual Taub-NUT solution presented in Eq.~\eqref{fnut} for the sake of simplicity. Notwithstanding, outside of the Taub-NUT class there is another configuration with a bolt that interests us and we devote the following section to it.

\subsection{Eguchi-Hanson\label{sec:Eguchi-Hanson}}

Having reviewed a known nonlinearly charged gravitational instanton, we are now in a better position to present a new one. Our objective in this section is to generalize the Eguchi-Hanson solution to the Einstein-Maxwell equations found in Ref.~\cite{Eguchi:1978gw}. That configuration is strictly Euclidean, that is, it has no Lorentzian analog. Its underlying topology is that of the sphere's tangent bundle and so its asymptotic boundary is a lens space. Geometrically, the background is described by a $U(1)$ fibration over a $2$-dimensional K\"ahler manifold. Thus, we consider an Eguchi-Hanson type line element given by
\begin{align}\label{metricansatz}
    \diff{s^2} = \frac{r^2f(r)}{4}\left(\diff{\psi} + \mathcal{B}_{(k)} \right)^2 + \frac{\diff{r^2}}{f(r)} + \frac{r^2}{4}\diff{\Sigma^2_{(k)}}\,,
\end{align}
where $\mathcal{B}_{(k)}$ and $\diff{\Sigma^2_{(k)}}$ are defined in Table~\ref{Table}. Once more, it is fruitful to assume that the gauge field is proportional to the K\"ahler one-form potential, namely,
\begin{align}
    A \equiv A_\mu\diff{x^\mu} = \alpha(r)\left(\diff{\psi} + \mathcal{B}_{(k)} \right)\,.
\end{align}

The linearly independent differential equations obtained from the ModMax and Einstein field equations are
\begin{align}
  \label{eomatEH}
  0  &= \alpha'' +  \frac{\alpha'}{r} - \frac{4 \alpha e^{-2\gamma}}{r^2}\,, \\
  \label{eomgttEH}
  0 &= \kappa \left[rf'+4\left(f-k\right)+\Lambda r^2\right] - \alpha'^2e^{\gamma} + \frac{4\alpha^2e^{-\gamma}}{r^2}\,,
\end{align}
respectively. The solution of this system is obtained following a similar prescription as in the Taub-NUT case, giving
\begin{subequations}\label{solEH}
\begin{align}\label{fsolEH}
    f(r) &= k - \frac{a^4}{r^4} - \frac{8pqe^{-\gamma}}{\kappa r^2} - \frac{\Lambda r^2}{6}\,, \\ 
    \alpha(r) &= q\, r^{-2e^{-\gamma}} + p\, r^{2e^{-\gamma}}\,,
\end{align}
\end{subequations}
with $a$, $p$, and $q$ being integration constants. This solution has a bolt at the largest root of the polynomial $f(r_b)=0$, subject to the condition $r_b\in\mathbb{R}_{>0}$. 

Similar to the Taub-NUT case, when $k=1$, the metric~\eqref{metricansatz} can be written in terms of the left-invariant Maurer-Cartan forms of $SU(2)$ [cf. Eq.~\eqref{MCformsu2}]
\begin{align}\label{Eguchi-Hanson-ansatz}
\diff{s^2} = \frac{\diff{r^2}}{f(r)} + \frac{r^2}{4} \left( \sigma_1^2 + \sigma^2_2 + {f(r)} \sigma_3^2\right)\,.
\end{align}
Notice that the hypersurfaces of constant radial coordinate are conformally equivalent to the Berger's sphere. Although this solution has a bolt, the Weyl tensor of this space is globally (anti-)self dual only if $\Lambda=0$ for arbitrary values of $p$ and $q$. This is different from the Taub-Bolt solution with ModMax electrodynamics, which is asymptotically (anti-)self dual even in presence of a nonvanishing cosmological constant~\cite{Flores-Alfonso:2020nnd}. Nevertheless, the gauge field is (anti-)self dual in the sense of Eq.~\eqref{selfdualmodmax} if and only if $p=0$. Thus, we shall focus on the case $p=\Lambda=0$ in Eq.~\eqref{solEH} from hereon, where the bolt radius becomes $r_b=a$. In that case, the range of coordinates is $a\leq r<\infty$, $0\leq\vartheta\leq\pi$, and $0\leq\varphi\leq2\pi$. On the other hand, the absence of conical singularities at the bolt implies that $0\leq\psi\leq2\pi$. Therefore, the degenerate hypersurface $r=a$ is topologically $\mathbb{S}^3/\mathbb{Z}^2=\mathbb{RP}^3$, while the asymptotic behavior is $\mathbb{R}^4/\mathbb{Z}^2$ being asymptotically locally Euclidean.

Before closing this section, let us recall that the Eguchi-Hanson space has been used to solve orbifold singularities of Calabi-Yau spaces in string theory~\cite{Polchinski:1998rr}. It has also been used as a seed metric to study gravitational solitons possessing nontrivial topology on the hypersurfaces of constant time~\cite{Clarkson:2005qx,Clarkson:2006zk,Chng:2006gh,Wong:2011aa,Hendi:2012zg,Durgut:2022xzw}. Additionally, higher-curvature corrections to the Eguchi-Hanson metric have been studied in Refs.~\cite{Corral:2021xsu,Corral:2022udb} and the backreaction of conformally coupled scalar fields was considered in Ref.~\cite{Barrientos:2022yoz}. Thus, future applications of the present configuration are plausible. 

In the next section, we study how the nonlinearly charged gravitational instantons presented here contribute to the axial anomaly of Dirac spinors. We compute the Dirac index explicitly, taking care of the boundary contributions.  

\section{Index theorem and Axial anomaly\label{sec:index}}

The Atiyah-Singer index theorem for the Dirac operator $\slashed{D}\equiv \gamma^a E^\mu_a D_\mu$ relates the number of positive and negative chiral spinors labeled by $N_\pm$ with topological invariants of the Pontryagin class~\cite{Atiyah:1968mp}.\footnote{Here, we denote by $E^\mu_a$ the inverse vierbein of $e^a_\mu$ which itself is defined through $g_{\mu\nu}=e^a_\mu e^b_\nu \delta_{ab}$, and it satisfies $e^a_\mu E^\mu_b = \delta^a_b$ and $e^a_\mu E^\nu_a=\delta^\nu_\mu$.} If the index of the Dirac operator is different from zero, one concludes that the nontrivial topology of the gauge connection induces a difference between the number chiral spinors. The aim of this section is to see whether the gravitational instantons studied in Sec.~\ref{sec:instantons} induce the axial anomaly or not. 

For noncompact manifolds, the Atiyah-Singer theorem is modified by boundary integrals of the Chern-Simons form and nonlocal terms related to the spectrum of the Dirac operator~\cite{APS-eta}. Moreover, if the spinors are charged under some gauge group, the index theorem receives additional contributions coming from their second Pontryagin class. For example, in four dimensions, the index theorem of the Dirac operator coupled to $U(1)$ gauge fields, i.e., $\slashed{D}=\gamma^c E^\mu_c (\partial_\mu + \frac{1}{4}\omega^{ab}{}_{\mu}\gamma_{ab} + iA_\mu)$, yields~\cite{Eguchi:1980jx,Franchetti:2017ftp}
\begin{widetext}
\begin{align}\notag
  N_+ - N_- &= -\frac{1}{24}\left[\frac{1}{32\pi^2} \int_{\mathcal{M}} \diff{}^4x\,\sqrt{|g|}\, \varepsilon_{\mu\nu \lambda\rho }R_{\sigma\tau}^{\lambda\rho}R^{\sigma\tau\mu\nu } - \frac{1}{4\pi^2}\int_{\partial\mathcal{M}}   \diff{^3x}\sqrt{|h|}\;n^\mu\varepsilon_{\mu\nu\lambda\rho}K^{\sigma\nu}\nabla^\lambda K^{\rho}_{\sigma}\right] \\
    \label{Diracindex}
    & + \frac{1}{32\pi^2}\int_{\mathcal{M}}\diff{^4x}\sqrt{|g|}\,\varepsilon_{\mu\nu\lambda\rho}F^{\mu\nu}F^{\lambda\rho} -\frac{1}{2}\left[\eta_D\left(\partial\mathcal{M} \right) + h_D\left(\partial\mathcal{M}\right) \right]\,,
\end{align}
\end{widetext}
where $\omega^{ab}=\omega^{ab}{}_\mu\,\diff{x^\mu}$ is the spin connection $1$-form, $\gamma_{ab}\equiv\gamma_{[a}\gamma_{b]}$ are the generators of the Lorentz group in the spinorial representation, $h_{\mu\nu}=g_{\mu\nu}-n_\mu n_\nu$ is the induced metric at the boundary $\partial\mathcal{M}$ with space-like unit normal $n^\mu$ and we have defined the extrinsic curvature as $K_{\mu\nu}=h^\lambda_\mu h^\rho_\nu \nabla_\lambda n_\rho$. Here, $\eta_D(\partial\mathcal{M})$ is the Atiyah-Patodi-Singer invariant that measures the difference between positive and negative eigenvalues of the Dirac operator on $\partial\mathcal{M}$ while $h_D(\partial\mathcal{M})$ captures the zero eigenvalues thereof~\cite{APS-eta}. The former has been recently interpreted as the axial charge of physical states~\cite{Kobayashi:2021jbn} and it can be obtained by performing the analytic continuation of the meromorphic function $\eta_D(s)$ to $s=0$, namely $\eta_D(\partial\mathcal{M}) = \eta_D(s)|_{s=0}$, where
\begin{align}\label{etas}
    \eta_D(s) = \sum_{\lambda\neq0}\text{sign}(\lambda)|\lambda|^{-s}\,,
\end{align}
with $\lambda$ being  the eigenvalues of the Dirac operator $\slashed{D}$ evaluated on the tangential components of $\partial\mathcal{M}$. The integral representation of Eq.~\eqref{etas} is given by
\begin{align}
    \eta_D(s) = \frac{1}{\Gamma\left(\frac{s+1}{2}\right)} \int_0^\infty\diff{x}\;x^{\frac{s-1}{2}}\,\text{tr}\left(\slashed{D}\,e^{-x\slashed{D}^2} \right)\,.
\end{align}
Then, if $\text{Index}(\slashed{D})=0$, there is no axial anomaly whatsoever. This is the case, for instance, of Dirac fermions on the (anti-)self-dual Taub-NUT background of general relativity~\cite{Eguchi:1977iu}. Nevertheless, it was shown by Pope that this is no longer true in presence of (anti-)self-dual Maxwell fields~\cite{Pope:1978zx,Pope:1981jx}. This implies that the axial symmetry is broken at the quantum level for charged Dirac spinors on the (anti-)self-dual Taub-NUT space. We will see here that this is the case in presence of nonlinear electrodynamics as well.

Equivalently, the Atiyah-Patodi-Singer $\eta_D$-invariant can be computed geometrically through the Hitchin formula~\cite{HITCHIN19741}. This can be done by writing the induced boundary metric on a hypersurface of constant radial coordinate, say $r=r_0$, conformally as the Berger's sphere, namely,
\begin{align}\label{bdymetric}
    \diff{s^2} &= \Omega(r_0)\left[\sigma_1^2 + \sigma_2^2 + \mu^2(r_0)\sigma_3^2 \right]\,,
\end{align}
where $\Omega(r_0)$ and $\mu(r_0)$ are positive definite functions of $r_0$.
It is worth mentioning that $\eta_D(s)$ is invariant under a constant rescaling of the eigenvalues $\lambda$, as it can be seen explicitly from its definition in Eq.~\eqref{etas}. Since the Dirac operator scales uniformly under conformal transformations of the boundary metric~\eqref{bdymetric}, one can set $\Omega(r_0)=1$ without loss of generality~\cite{Bakas:2011nq}. Then, in vacuum, the Hitchin's formula establishes that the $\eta_D$-invariant can be computed from the boundary metric~\eqref{bdymetric} through~\cite{HITCHIN19741}
\begin{align}
    \eta_D\left(\partial\mathcal{M} \right) + h_D\left(\partial\mathcal{M}\right) = \frac{1}{6}\left(1-\mu^2_0 \right)^2\,,
\end{align}
where $\mu_0=\mu(r_0)$. This provides a convenient prescription to obtain the axial anomaly of Dirac spinors from a geometrical viewpoint.

To compute the $\eta_D$-invariant in presence of ModMax fields, we consider the induced boundary metric~\eqref{bdymetric} over a hypersurface of constant radius. Since the Dirac operator is invariant under a constant rescaling of the eigenvalues, without loss of generality we can choose a dreibein basis $1$-form $e^a=e^a_\mu\diff{x^\mu}$ as
\begin{align}
e^1 &= \sigma_1\,, & e^2 &= \sigma_2\,, & e^3 &= \mu_0\,\sigma_3\,.
\end{align}
The dual vector basis $E_a=E^\mu_a\partial_\mu$ can be computed from the orthogonality condition with the dreibein one-forms, that is, $\langle e^a,E_b\rangle=\delta^a_b$. Then, we obtain 
\begin{align}
    E_1 &= \Sigma_1\,, & E_2 &= \Sigma_2\,, & E_3 &= \mu^{-1}_0\Sigma_3\,,
\end{align}
where $\Sigma_i=\Sigma_i^\mu\partial_\mu$ with $i=1,2,3$ are the dual vector basis to the left-invariant Maurer-Cartan one-forms of $SU(2)$~\eqref{MCformsu2}; that is,
\begin{subequations}
\begin{align}
    \Sigma_{1}&=-\cot\vartheta\sin\psi\,\partial_{\psi}+\cos\psi\,\partial_{\vartheta}+\frac{\sin\psi}{\sin\vartheta}\,\partial_{\varphi}\,, \\
    \Sigma_{2}&=-\cot\vartheta\cos\psi\,\partial_{\psi}-\sin\psi\,\partial_{\vartheta}+\frac{\cos\psi}{\sin\vartheta}\,\partial_{\varphi}\,, \\
    \Sigma_3 &=\partial_{\psi}\,.
\end{align}
\end{subequations}
On the other hand, the nontrivial components of the spin connection can be obtained by solving the torsion-free condition $\diff{e^a}+\omega^{a}{}_b\wedge e^b=0$, giving
\begin{align}
\omega^{12} &=\left(\frac{\mu^2_0-2}{2\mu_0}\right) e^3\,, & \omega^{13} &= \frac{\mu_0\,e^2}{2}\,, & \omega^{23} &= -\frac{\mu_0\,e^1}{2}\,.
\end{align}
For the Dirac matrices, we choose the representation in a co-dimension $1$ hypersurface of constant radius as
\begin{align}
    \gamma^{1}&=\begin{pmatrix}0 & 1\\
1 & 0
\end{pmatrix}, & \gamma^{2}&=\begin{pmatrix}0 & -i\\
i & 0
\end{pmatrix}, & \gamma^{3}&=\begin{pmatrix}1 & 0\\
0 & -1
\end{pmatrix}\,.
\end{align}
Additionally, we denote the $U(1)$ connection $1$-form evaluated at $r=r_0$ by $A=b(r_0)\sigma_3\equiv b_0\sigma_3$.\footnote{In the Taub-NUT case, this definition implies that $b(r)=2n\alpha(r)$ provided the coordinate transformation $\psi\to\frac{\psi}{2n}$, while for Eguchi-Hanson we have $b(r)=\alpha(r)$.} Then, defining the self-adjoint operator $K_i = i\Sigma_i$ and the ladder operator $K_\pm=K_1 \pm iK_2$, one can show that they satisfy the Lie algebra of angular momentum, i.e. $\left[K_i,K_j\right]=i\epsilon_{ijk}K^k$, $\left[K_3,K_\pm\right]=\pm K_\pm$, and $\left[K_+,K_-\right]=K_3$. Hence, the Dirac operator acting on a two-component Dirac spinor expressed in matrix notation can be written as
\begin{align}\notag 
    \slashed{D}\Psi &= \begin{pmatrix}\mu_0^{-1}\left(K_{3}+b_0\right) & K_-\\
K_+ & -\mu^{-1}_0\left(K_{3}+b_0\right)
\end{pmatrix}\begin{pmatrix}\Psi_{1}\\
\Psi_{2}
\end{pmatrix} \\ 
&\quad +\frac{(\mu^2_0+2)}{4\mu_0}\begin{pmatrix}\Psi_{1}\\
\Psi_{2}
\end{pmatrix}=0\,.
\end{align}

In order to compute the spectrum of the Dirac operator, we follow Ref.~\cite{Pope:1981jx} and denote $|s\rangle\equiv \,_sY_{lm}(\vartheta,\varphi)$, where $_sY_{lm}(\vartheta,\varphi)$ stands for the spin-$s$ spherical harmonics~\cite{Goldberg:1966uu}. Then, the self-adjoint operator acting on the state $|s\rangle$ gives  
\begin{align}
    K_\pm\, |s\rangle &= \sqrt{(l\mp s)(l\pm s+1)}\; |s\pm1\rangle\,, \\
    K_3\, |s\rangle &= s\,|s\rangle\,,
\end{align}
for $|s|\leq l$ and $|m|\leq l$. The eigenvectors of $\slashed{D}$ can be written as $\Psi_1 = c_1\,|s\rangle$ and $\Psi_2 = c_2\,|s+1\rangle$, where $c_1$ and $c_2$ are constants. Thus, for $-l\leq s\leq l-1$, the eigenvalues $\lambda$ of the Dirac operator $\slashed{D}$ are~\cite{Pope:1981jx}
\begin{align}\label{eigenvalues}
    \lambda = \frac{\mu_0}{2} \pm \mu_0^{-1}\sqrt{(2s+1+2b_0)^2+4\mu^2_0(l-s)(l+s+1)}\,,
\end{align}
where, for each value of $s$, these eigenvalues have a degeneracy of $d=2l+1$. Notice that there are two special cases: (i) if $s=l$, or (ii) if $s=-(l+1)$ where $c_2$ and $c_1$ vanish, respectively. Then, their eigenvalues become
\begin{subequations}\label{speciallambda}
\begin{align}
    \mbox{(i)}\;\;\;\;\; \lambda &= \frac{\mu_0}{2} \pm \mu_0^{-1}\left|2l+1+2b_0\right|\,,\\
    \mbox{(ii)}\;\;\;\;\; \lambda &= \frac{\mu_0}{2} \pm \mu_0^{-1}\left|2l+1-2b_0\right|\,.
\end{align}
\end{subequations}

We are interested in evaluating the $\eta_D$-invariant at the asymptotic boundary, that is, as $r_0\to\infty$. For the sake of simplicity, from hereon we focus on the case with vanishing cosmological constant or, equivalently, $\ell\to\infty$.  Then, for the (anti-)self-dual Taub-NUT (TN) and Eguchi-Hanson (EH) spaces, we have
\begin{align}
    &\mbox{TN:} & \lim_{r_0\to\infty}\mu(r_0) &= 0 & &\wedge & \lim_{r_0\to\infty}b(r_0) &= q\,,\\
    &\mbox{EH:} & \lim_{r_0\to\infty}\mu(r_0) &= 1 & &\wedge & \lim_{r_0\to\infty}b(r_0) &= 0\,,
\end{align}
respectively. Notice that the ModMax field approaches to its Maxwellian counterpart in the charged Euclidean Taub-NUT metric, as mentioned in Sec.~\ref{sec:instantons}. On the other hand, in the case of the Eguchi-Hanson instanton, the ModMax field decays sufficiently fast toward the asymptotic boundary. Therefore, the $\eta_D$-invariant of the latter coincides with the one computed in Ref.~\cite{Eguchi:1978gw}, giving 
\begin{align}\label{etaEH}
    \eta_D^{\rm (EH)}\left(\partial\mathcal{M}\right) + h_D^{\rm (EH)}\left(\partial\mathcal{M}\right) = \frac{1}{4}\,. 
\end{align}  
Although the boundary contributions to the Dirac index in Eq.~\eqref{Diracindex} vanishes for the charged Eguchi-Hanson solution of Sec.~\ref{sec:Eguchi-Hanson}, the bulk part gives
\begin{align}
\frac{1}{32\pi^2} \int_{\mathcal{M}} \diff{}^4x\,\sqrt{|g|}\, \varepsilon_{\mu\nu \lambda\rho }R_{\sigma\tau}^{\lambda\rho}R^{\sigma\tau\mu\nu } &= -3 \\
\frac{1}{32\pi^2}\int_{\mathcal{M}}\diff{^4x}\sqrt{|g|}\,\varepsilon_{\mu\nu\lambda\rho}F^{\mu\nu}F^{\lambda\rho} &= - \frac{q^2}{a^{4e^{-\gamma}}}\,. \label{PontMaxEH}
\end{align}
In this case, the topology of the bolt introduces a screening factor on the topological charge~\eqref{PontMaxEH} that can be controlled with the nonlinear parameter $\gamma$. Indeed, in the limit $\gamma\to\infty$, the latter becomes equal to $-q^2$. This point in the parameter space is of particular interest as, in the manifest duality-symmetric formulation of Ref.~\cite{Avetisyan:2021heg}, the ModMax and Bia{\l}ynicki-Birula electrodynamics are part of the same one-parameter family. Moreover, that reformulation suggests that the Bia{\l}ynicki-Birula electrodynamics roughly corresponds to the ModMax limiting case $\gamma\to\infty$.

In summary, for a finite $\gamma$, we find that the Dirac index for the charged Eguchi-Hanson solution of Einstein-Modmax theory is 
\begin{align}\label{anomalyEH}
    N_+-N_- =  -\frac{q^2}{a^{4e^{-\gamma}}}\,,
\end{align}
which is nonvanishing for $q\neq0$ with $\gamma\in\mathbb{R}$. Thus, we conclude that this particular instantonic configuration does contribute to the axial anomaly.
\pagebreak

The computation of the $\eta_D$-invariant in the case of the (anti-)self-dual Taub-NUT instanton is more subtle, since the eigenvalues~\eqref{eigenvalues} are divergent as $r_0\to\infty$. Nevertheless, one can exploit the fact that the $\eta_D(s)|_{s=0}$ is invariant under the global rescaling $\lambda\to\mu_0\lambda$ to obtain finite eigenvalues. In order to do so, we first perform the rescaling of the eigenvalues and then take the limit $r_0\to\infty$. This procedure yields~\cite{Pope:1981jx}
\begin{align}\label{eigenvaluesrescaled}
    \lambda = \pm|2s+1+2b_0|\,.
\end{align}
The symmetry between the positive and negative eigenvalues derived from this equation implies that their sum will not contribute to the $\eta_D$-invariant. However, this is not the case if $s=l$ or if $s=-(l+1)$, which leads to the particular values in Eq.~\eqref{speciallambda}. Thus, after rescaling the eigenvalues as $\lambda\to\mu_0\lambda$ and taking the limit $r_0\to\infty$, one obtains
\begin{subequations}\label{particularlambda}
\begin{align}\label{lambda1}
    \mbox{(i)}\;\;\;\;\; \lambda &= 2l+1+2b_0\,,\\
    \label{lambda2}
    \mbox{(ii)}\;\;\;\;\; \lambda &= 2l+1-2b_0\,,
\end{align}
\end{subequations}
both with degeneracy $d=2l+1$. These exceptional cases are the main contribution to the $\eta_D$-invariant. To see this, let us focus on the case when $b_0$ is an integer with $b_0>0$ without loss of generality; the case for noninteger $b_0$ in Einstein-Maxwell theory has been studied in Ref.~\cite{Pope:1981jx}. Then, the eigenvalue~\eqref{lambda1} is positive definite, while Eq.~\eqref{lambda2} could take negative values if $2l+1<b_0$. Thus, considering the contributions coming from the positive and negative eigenvalues of the Dirac operator to the $\eta_D$-invariant in Eq.~\eqref{etas}, the latter can be expressed as
\begin{widetext}
\begin{align}\notag
    \eta_D^{\rm (TN)}\left(\partial\mathcal{M} \right) &= \lim_{s\to0}\left[ \sum_{l=0}^\infty \left(2l+1\right)\left(2l+1+2b_0\right)^{-s} - \sum_{l=0}^{2b_0-1}\left(2l+1\right)\left(2l+1-2b_0\right)^{-s} + \sum_{l=2b_0}^{\infty}\left(2l+1\right)\left(2l+1-2b_0\right)^{-s}\right] \\
    &= -\frac{1}{6} - 2q\,, \label{etaTN}
\end{align}    
\end{widetext}
where we have used the fact that $h_D^{\rm (TN)}(\partial\mathcal{M})=0$ and the definition of the Riemann zeta function $\zeta(s)=\sum_{p=1}^{\infty}p^{-s}$. Additionally, some of its particular values has been taken into account, e.g. $\zeta(-1)=-1/12$ and $\zeta(0)=-1/2$. Then, inserting Eq.~\eqref{etaTN} into the Dirac index~\eqref{Diracindex} and using the bulk integrals
\begin{align}
\frac{1}{32\pi^2} \int_{\mathcal{M}} \diff{}^4x\,\sqrt{|g|}\, \varepsilon_{\mu\nu \lambda\rho }R_{\sigma\tau}^{\lambda\rho}R^{\sigma\tau\mu\nu } = 2\,, \\
\frac{1}{32\pi^2}\int_{\mathcal{M}}\diff{^4x}\sqrt{|g|}\,\varepsilon_{\mu\nu\lambda\rho}F^{\mu\nu}F^{\lambda\rho} = 2q^2 \,,\label{U(1)CPindexTN}
\end{align}
together with the fact that the boundary integral of the Chern-Simons form vanishes for the (anti-)self-dual Taub-NUT instanton, we find
\begin{align}\label{anomalyTN}
    N_+-N_- = 2q\left(q+\frac{1}{2}\right)\,.
\end{align}
Thus, we conclude that, in the case of the (anti-)self-dual Taub-NUT solution of Einstein-ModMax theory, the axial anomaly is sourced by the electric charge of the ModMax field. Indeed, its value does not depend on the new parameter $\gamma$ that measures the nonlinearity of the electromagnetic fields and it coincides with the Maxwell case~\cite{Pope:1981jx}. This is related to the fact that the asymptotic behavior of the ModMax field does not modify the $\eta_D$-invariant in comparison to the linear Maxwell case. Moreover, the $U(1)$ Chern-Pontryagin index~\eqref{U(1)CPindexTN} has no dependence on the $\gamma$-parameter whatsoever, in contrast to the Eguchi-Hanson case. This is related to the fact that the latter is endowed with a bolt and we expect that it will not be necessarily the case in the Taub-Bolt instanton either that we aim to study in the future. 

\section{Discussion\label{sec:discussion}}

In this work, we have shown that gravitational instantons sourced by nonlinear conformal electrodynamics induce the axial anomaly when coupled to Einstein gravity. For doing so, we first review the Taub-NUT solution found in this theory in Refs.~\cite{BallonBordo:2020jtw,Flores-Alfonso:2020nnd}. In the asymptotic regime, we verify that the solution of the ModMax field equations approaches its Maxwellian counterpart in this background. Moreover, the former is continuously connected to the latter as $\gamma\to0$; this parameter measures the degree of nonlinearity of the $U(1)$ gauge fields. Then, we obtain a novel Eguchi-Hanson instantonic solution in Einstein-ModMax theory. Similar to the Taub-NUT case, we find that the $\gamma$-parameter screens the electric and magnetic charge of the dyonic nonlinear $U(1)$ fields. By demanding (anti-)self duality and the absence of conical singularities, we find a nonlinearly charged, geodesically complete lens space that is asymptotically locally Euclidean.

Then, by using the Atiyah-Patodi-Singer theorem, we compute the index of the Dirac operator for nonlinearly charged spinors. Although there is no contribution of the ModMax field to the $\eta_D$-invariant in the Eguchi-Hanson instanton, its nontrivial topology at the bolt induces a nonvanishing $U(1)$ Chern-Pontryagin index that contributes to the anomaly, detecting the presence of the nonlinear theory at the level of the topological charge. This effect of the nonlinear electrodynamics is somewhat similar to charge screening. Additionally, in the (anti-)self-dual Taub-NUT case, the gauge potential contributes nontrivially to the $\eta_D$-invariant while the Chern-Pontryagin index is trivial. We compute the former by performing the analytic continuation of the meromorphic function~\eqref{etas} to $s=0$ and evaluating the difference between the number of positive and negative eigenvalues of the Dirac operator. Then, we conclude that both instantons induce an excess of positive chiral spinors in comparison with the negative ones [cf. Eqs.~\eqref{anomalyEH} and~\eqref{anomalyTN}]. 


Interesting questions remain open. First, studying harmonic forms and spinors for the nonlinearly charged Taub-bolt is certainly of interest, similar to the analysis performed in Ref.~\cite{Franchetti:2018woq}. Additionally, even though the computation in presence of a negative cosmological constant is more involved, it could provide important insights regarding the role of nonlinear $U(1)$ fields in the axial anomaly from a conformal field theory viewpoint~\cite{Maldacena:1997re,Gubser:1998bc,Witten:1998qj}. Some advances in this direction have been done in Ref.~\cite{Franchetti:2022tga} where the space of $SU(2)$-invariant harmonic 2-forms was determined for Taub-NUT-AdS solution in Einstein-Maxwell theory. Finally, it is worth analyzing whether quantum effects would break the conformal symmetry of ModMax fields. This is of interest to us and we postpone a deeper study of this point to the future.

\begin{acknowledgments}
We thank Giorgos Anastasiou, Ignacio J. Araya, Eloy Ay\'on-Beato, Fabrizio Canfora, Gast\'on Giribet, Olivera Mi\v{s}kovi\'c, Rodrigo Olea, Francisco Rojas, and Omar Valdivia for insightful comments and discussions. The work of F.C-M. is supported by Agencia Nacional de Investigaci\'{o}n y Desarrollo (ANID) through FONDECYT No. 1210876. C.C. is partially supported by FONDECYT grants No.~11200025 and~1210500. D.F-A. is supported by ANID under FONDECYT grant No. 3220083. L.S. is supported by Beca de Doctorado Nacional ANID No. 21221813.
\end{acknowledgments}

\bibliography{References}

\end{document}